\documentclass[prl,aps,twocolumn,showpacs,superscriptaddress]{revtex4}

\usepackage{latexsym,epsfig,amsfonts,amssymb,amsmath,graphicx}

\newcommand{\av}[1]{\langle #1\rangle}

\allowdisplaybreaks

\begin{document}

\title{High field fractional quantum Hall effect in optical lattices}

\author{R.N. \surname{Palmer}}%
\affiliation{Clarendon Laboratory, University of Oxford, Parks Road, Oxford OX1 3PU, UK}%
\author{D. \surname{Jaksch}}%
\affiliation{Clarendon Laboratory, University of Oxford, Parks Road, Oxford OX1 3PU, UK}%

\date{\today}

\begin{abstract}
We consider interacting bosonic atoms in an optical lattice subject
to a large simulated magnetic field. We develop a model similar to a
bilayer fractional quantum Hall system valid near simple rational
numbers of magnetic flux quanta per lattice cell. Then we calculate
its ground state, magnetic lengths, fractional fillings, and find
unexpected sign changes in the Hall current. Finally we study
methods for detecting these novel features via shot noise and Hall
current measurements.
\end{abstract}


\maketitle

The achievement of strongly correlated atomic systems \cite{BHM-exp}
which are accurately described by simple condensed matter physics
(CMP) models \cite{BHM-theo} has opened many new perspectives for
research on strongly correlated systems. Such atomic ensembles share
many properties with conventional CMP systems. Several of their
features, however, distinguish them from CMP setups like e.g.~their
extremely long coherence times and the ability to adjust Hamiltonian
parameters over a wide range and in times short compared to the
coherence time. These properties enable experiments of controlled
coherent quantum dynamics \cite{BHM-exp, Esslinger}, and the
exploration of otherwise unaccessible parameter regimes. Due to
their highly non-linear behavior these studies may lead to striking
new features.

In this paper we consider such a novel situation arising if
interacting ultracold bosonic atoms in a two dimensional optical
lattice are subjected to a simulated huge magnetic field. A variety
of proposals for realizing magnetic fields in optical lattices
exist. One can e.g.~rotate the lattice as suggested in
\cite{bec-anyons,rotating-lattice}, use laser-induced hopping as
proposed in \cite{butterfly}, or employ an oscillating quadrupole
potential as shown in \cite{laughlin-lattice}. All of these methods
allow the creation of arbitrarily large fields with flux quanta per
lattice cell $\alpha$ between $0 \leq \alpha \leq 1$ and can be
realized with current or near future experimental technology.
Optical lattice setups are virtually free of imperfections and can
be cooled to almost zero temperature \cite{BHM-exp}. Furthermore,
their large degree of flexibility \cite{lattice-review} allows for
the introduction of well controlled amounts of disorder
\cite{disorder}. This is in contrast to most CMP systems where only
values $\alpha \ll 1$ are believed to be accessible and disorder is
difficult to determine and control. Also, conventional CMP systems
have fermions with a screened $1/r$ Coulomb interaction while we
consider bosonic atoms interacting via a contact potential. Finally,
dynamically adjustable smooth trapping potentials are used to
confine atoms in an optical lattice to a certain region in space
\cite{lattice-review} while CMP systems are often bounded by sharp
edges.

For a single particle in a lattice placed into a huge magnetic field
with $\alpha \sim 1$ a fractal energy band structure was predicted
by Hofstadter \cite{Hof}. However, this regime was not thought to be
experimentally accessible until recently and is thus not explored in
detail. Many studies on interaction effects in magnetic fields
centered around the fractional quantum Hall (FQH) effect for $\alpha
\ll 1$ \cite{fqh-text} where it is essentially unaffected by the
presence of a lattice. As $\alpha \sim 1$ the discrete periodic
structure of the lattice becomes important and we will find novel
effects arising from the presence of the lattice. We will show that
the system is described by a bilayer FQH like model near simple
rational values $\alpha_c \equiv l/n$ where $n$ is a small integer.
The ground state of this system is an FQH state with a mean filling
factor $\tilde \nu = n/(n+1)$ and a modified magnetic length. We
will find striking features of the associated Hall current which
changes its sign at $\alpha = \alpha_c$.

We consider ultracold bosonic atoms in the lowest motional band of a
3D optical lattice with lattice period $d$. The motion of the atoms
along the $z$-direction is suppressed by high optical potential
barriers \cite{lattice-review}. The system is thus decoupled into 2D
planes and we assume that a large effective magnetic field along the
$z$-axis is simulated by one of the methods introduced in
\cite{rotating-lattice,butterfly,laughlin-lattice}. The Hamiltonian
in one of the planes is given by
\begin{eqnarray}
H&=&-J\sum\limits_{p,q} (e^{2\pi i\alpha q}
a^\dagger_{p,q}a_{p-1,q}+a^\dagger_{p,q}a_{p,q-1} +{\rm h.c.})\nonumber\\
&&\quad+V(p,q)a^\dagger_{p,q}a_{p,q}+
\frac{U}{2}a^\dagger_{p,q}a^\dagger_{p,q}a_{p,q}a_{p,q}.
\end{eqnarray}
Here $a^\dagger_{p,q}$ ($a_{p,q}$) are bosonic creation
(annihilation) operators, fulfilling standard bosonic commutation
relations, for particles at a lattice site labeled by its position
in the $xy$-plane $(p,q)$. The onsite interaction strength is $U$
and $J$ is the hopping amplitude. The parameter $\alpha$ is a
measure for the strength of the simulated magnetic field
\cite{butterfly}. The phase shift from hopping around one lattice
cell is $2\pi\alpha$, and since $\alpha$ is only defined mod 1 we
restrict $-1/2\leq\alpha <1/2$. Finally, $V(p,q)$ is a trapping
potential varying slowly on the length scale $d$. We will mostly
deal with linear geometries $V(p,q)\equiv V(q)$, where the
$x$-dependence of the atomic wavefunctions are plane waves, and thus
work in Landau gauge. All parameters in the Hamiltonian can
dynamically be varied via lasers \cite{lattice-review}.


We first discuss the limit $|\alpha| \ll 1$, in which the
lengthscale of the wavefunctions is much larger than $d$ and we can
hence use a continuum approximation. We approximate a one particle
state $|\psi\rangle= \sum_{p,q} \psi (p,q) a^\dagger_{p,q}|{\rm
vac}\rangle$ with $|{\rm vac}\rangle$ the vacuum state by a
continuous wave function $\phi(pd,qd)=\psi(p,q)/d^2$. The
Hamiltonian acting on $\phi$ is
\begin{equation}
H_0=-\frac{1}{2m}\left[\frac{\partial^2}{\partial y^2}+
\left(2m\Omega y-i\frac{\partial}{\partial
x}\right)^2\right]+V(x,y)-\frac{2}{md^2},\label{eq:cont-1par}
\end{equation}
where we have defined the effective mass $m=1/2Jd^2$ and the
cyclotron frequency $\Omega=\pi \alpha/m d^2$. The last term has no
dynamical effect and is left out in the following. For interacting
particles at density $\rho$, and interparticle spacing
$\rho^{-1/2}\gg d$, the continuum approximation is
\begin{equation}
H\approx\sum\limits_i H_0(x_i,y_i)+\frac{u}{2} \sum\limits_{i\neq
j}\delta(x_i-x_j)\delta(y_i-y_j), \label{eq:cont-manypart}
\end{equation}
where $(x_i,y_i)$ are the coordinates of particle $i$ and $u=Ud^2$.
By analogy with the solid state FQH we define the filling factor
$\nu=\rho\pi/m\Omega$. The 2D Hamiltonian
Eq.~(\ref{eq:cont-manypart}) has been studied e.g.~in the context of
rotating ultracold atomic gases for the case of a weak trap
\cite{bec-anyons,bec-lll-expt,laughlin-lattice,laughlin-rotn} and it
is well established that for an isotropic trap $V(x,y)=m\omega^2
(x^2+y^2)/2$ the $\nu=1/2$ Laughlin state with a magnetic length
$\ell$ given by $\ell_c=(m^2\Omega^2+ m^2\omega^2)^{-1/4}$ is an
{\it exact} eigenstate. Its energy is $N(\Omega^2+ \omega^2)^{1/2}/2
+N(N-1)\Delta E$, where $\Delta E\equiv (\Omega^2+\omega^2)^{1/2}
-\Omega\approx\omega^2/2\Omega$ and $N$ is the number of particles.
It is the ground state of the system for $\omega \ll \Omega,
um\Omega$. Since this state already has exactly zero interaction
energy, the rest of the bosonic Laughlin sequence
($\nu=1/4,1/6,\ldots$) does not occur. It is also possible to
construct non-circular Laughlin-like states for other trap
geometries. In general these are only exact eigenstates for zero
trap strength, but as they are always lowest Landau level (LLL) and
zero interaction energy, they are reasonable trial states for weak
traps. For instance, in linear geometry with a trap $V(y)=m\omega^2
y^2/2$ the variational energy is minimised for a Laughlin like state
with magnetic length $\ell_l=(m^2\Omega^2 +m^2\omega^2/4)^{-1/4}$.

These Laughlin states cannot remain the ground states if $N$ gets
too large as for $N\gtrsim u/4\pi \ell^2\Delta E$ the trap potential
energy $2N\Delta E$ to add a particle to the edge of the Laughlin
state exceeds the interaction energy $\sim u/2\pi\ell^2$ required to
add it in the centre of the trap. For larger $N$, denser states are
more favourable, and it is believed \cite{readrez-steps} that a
series of incompressible states with half-integer filling
$\nu=1,3/2,2,\ldots$, well approximated by symmetrized products of
Laughlin states called Read-Rezayi states \cite{readrez-def}, forms.
For $\nu \gtrsim 6$ a transition to a compressible vortex lattice is
predicted \cite{readrez-steps}.


\begin{figure}
\includegraphics[width=7cm]{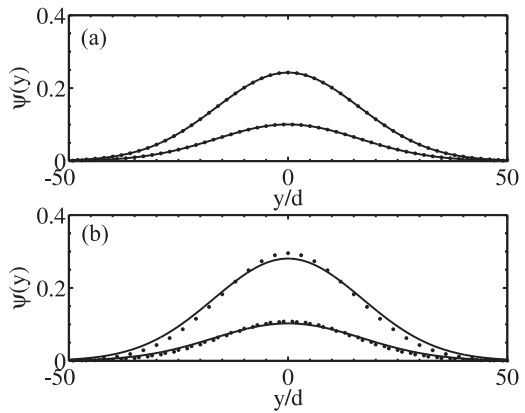}
\caption{Wavefunction for (a) $\alpha=1/2$ and (b) $\alpha=1/3$, in
linear geometry with trap $V(y)=m\omega^2 y^2/2$ and
$\omega=10^{-3}\pi/md^2$. The dots show the numerical wavefunction
and the solid curves the analytic approximations for $y/d\equiv 0$
mod $n$ (upper curve) and $y/d\neq 0$ mod $n$ (lower curve).}
\label{fig:2cpt-wavefn}
\end{figure}

We now turn to the case of larger values of $\alpha$ which can be
reached in optical lattice setups \cite{butterfly}. We restrict our
considerations to values of $\alpha$ close to simple rationals
$\alpha_c\equiv l/n$, where $l,n$ are small integers. For a single
particle we find from numerical calculation shown in
Fig.~\ref{fig:2cpt-wavefn} that when $\alpha\approx \alpha_c$ the
single particle ground state wavefunction has an approximate
$n$-site periodicity, suggesting the representation
$\psi(np+i,nq+j)=d^2\chi_k(npd,nqd){\bf B}^{(k)}_{ij}$ where
$\chi_k$ is a continuous function and ${\bf B}^{(k)}$ an $n\times n$
matrix. We find by expansion about $\alpha_c$ that there are $n$
linearly independent matrices ${\bf B}^{(k)}$ with degenerate
energies, of the form ${\bf B}^{(k)}_{pq}=e^{2\pi ipkl/n}{\bf
v}_{q-k}$ where ${\bf v}$ is a fixed $n$-component vector for each
$l,n$; e.g.~for $\alpha_c=1/2$ we have ${\bf
v}=\left(\sqrt{1-1/\sqrt{2}}, \sqrt{1+1/\sqrt{2}}\right)$ for the
minimum energy eigenvector. The subscript $q-k$ wraps around mod
$n$, and the $\chi_k$ obey
\begin{equation}
-\frac{C}{2m}\left[\frac{\partial^2}{\partial y^2}+
\left(2m\tilde\Omega y-i\frac{\partial}{\partial
x}\right)^2\right]\chi_k+V(x,y)\chi_k =E\chi_k, \label{eq:chi-1par}
\end{equation}
where $\tilde\Omega\equiv(\alpha-\alpha_c)\pi/md^2$ and $C$ depends
only on $l,n$. This formula reduces to Eq.~(\ref{eq:cont-1par}) for
$\alpha_c=0/1$. The comparison in Fig.~\ref{fig:2cpt-wavefn} shows
excellent agreement of the wave functions for the cases
$\alpha_c=1/2$ and $\alpha_c=1/3$.

\begin{figure}
\includegraphics[width=7cm]{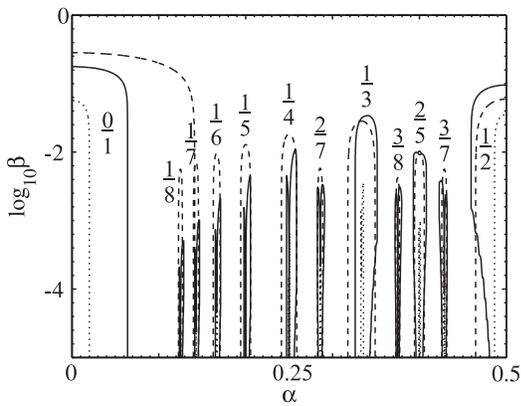}
\caption{Areas of validity of different fractions $\alpha_c$, in a
one-dimensional harmonic trap. The solid (dotted) curves show
regions with overlap between exact numerical wavefunctions and their
analytical approximations larger than $99\%$ ($99.9\%$) and
resulting energy difference smaller than $0.01/md^2$ ($0.001/md^2$).
The dashed lines are areas of validity obtained from
Eq.~(\ref{eq:chi-validrange}) taking $\gg$ to mean a ratio larger
than $50$.} \label{fig:overlap}
\end{figure}

For this procedure to be consistent, the magnetic length $\tilde
\ell$ on which $\chi_k$ varies must be large compared to the small
scale periodicity $nd$, but small enough that $m\tilde\Omega y d \ll
1$ for $y \sim \tilde\ell$. For a 1D harmonic trap
$\tilde\ell_l=(m^2\tilde\Omega^2+m^2 \omega^2/4C)^{-1/4}$ and thus
the analytic approximation is valid for
\begin{equation}
\frac{1}{n^4}\gg(\alpha-\alpha_c)^2+\frac{\beta^2}{4C}\gg
(\alpha-\alpha_c)^4,\label{eq:chi-validrange}
\end{equation}
where $\beta \equiv md^2\omega/\pi$ is the dimensionless trap
strength. In Fig.~\ref{fig:overlap} we compare approximate
analytical and exact numerical results for values of $n\leq 8$ as a
function of $\alpha$ and $\beta$. The plot shows large regions of
validity of our calculation around $\alpha_c=0$ (conventional FQH
region) but also around $\alpha_c=1/2$ and $\alpha_c=1/3$. As
expected for increasing $n$ the regions of validity get narrower and
are more easily destroyed by an external potential. The regions also
get narrower for decreasing values of $\beta$ with at the same time
increasing number of $n$'s fulfilling Eq.~(\ref{eq:chi-validrange})
at $\alpha=\alpha_c$. Our model is thus only valid in a finite
neighborhood of an $\alpha_c$ if a suitable trapping potential is
present.

We extend the Hamiltonian to the many particle case by including an
onsite interaction term into Eq.~(\ref{eq:chi-1par}) giving
\begin{eqnarray}
&&H\approx\int dx dy
\sum\limits_k\chi_k^\dagger\left\{\frac{C}{2m}\left[-\frac{\partial^2}{\partial
y^2}+\left(2m\tilde\Omega y-i\frac{\partial}{\partial
x}\right)^2\right]\right. \nonumber\\
&& \left.+V(x,y)\right\}\chi_k +u
\sum\limits_{k_1,k_2,k_3,k_4}G_{k_1,k_2,k_3,k_4}\chi_{k_1}^\dagger\chi_{k_2}^\dagger\chi_{k_3}\chi_{k_4},\label{eq:chi-manypar}
\end{eqnarray}
where $\chi_k(x,y)$ is now a bosonic field operator and the
constants $G_{k_1,k_2,k_3,k_4}\equiv\sum_j {\bf v}_{j-k_1}{\bf
v}_{j-k_2}{\bf v}_{j-k_3}{\bf v}_{j-k_4}/n$ if $k_1+k_2\equiv
k_3+k_4$ mod $n$ and 0 otherwise; e.g.~for $\alpha_c=1/2$,
$G_{1111}=G_{2222}=3/2$ and
$G_{1212}=G_{2121}=G_{1122}=G_{2211}=1/2$. Particles with the same
$k$ interact more strongly since their ${\bf v}^{(k)}$ are peaked on
the same sites.

For $\alpha_c=1/2$ this effective Hamiltonian is equivalent to a
bilayer FQH system with $\chi_\pm=\chi_1\pm i\chi_2$ being the two
"layers", and hence for weak potentials $V$ the highest density zero
interaction energy state is the 221 state. This bosonic analogue of
the fermion 331 state \cite{fqh-text} with magnetic length $\tilde
\ell_0=(m\tilde\Omega)^{-1/2}$ has filling factor $\tilde\nu\equiv
\rho\pi\tilde\ell_0^2=2/3$; it can be extended to larger $n$ with
$\tilde\nu=n/(n+1)$ by multiplication with $(z_i-z_j)^2$ if
particles $i,j$ are in the same "layer" and with $(z_i-z_j)$ if they
are not. We expect a density step profile eventually breaking down
into a vortex lattice, qualitatively similar to the small $\alpha$
regime, with the 221 state being the lowest step. As long as
Eq.~(\ref{eq:chi-manypar}) is a valid approximation, i.e.~the
particle spacing is much larger than $nd$, every step would occur at
fixed $\tilde\nu$ and not fixed $\nu$. From Fig.~\ref{fig:overlap},
the range of validity for $\alpha_c=1/2$ is $\alpha-\alpha_c\lesssim
0.04$ and hence for the $\tilde\nu=2/3$ state, lattice filling $\rho
d^2\lesssim 0.03$. For $U\lesssim J$ the energy gap is $E_g \sim
U\rho d^2$ which is about $10$ - $100$Hz for typical optical lattice
parameters consistent with \cite{laughlin-lattice} and will thus be
measurable with near future technology. Measurement of the stepped
density profile would confirm the existence of a series of
incompressible states near $\alpha=\alpha_c$ of densities
proportional to $\tilde\Omega$ but would not confirm that they are
FQH states. We will thus next investigate two methods feasible in
optical lattices to identify and characterize these states further.

\begin{figure}
\includegraphics[width=7cm]{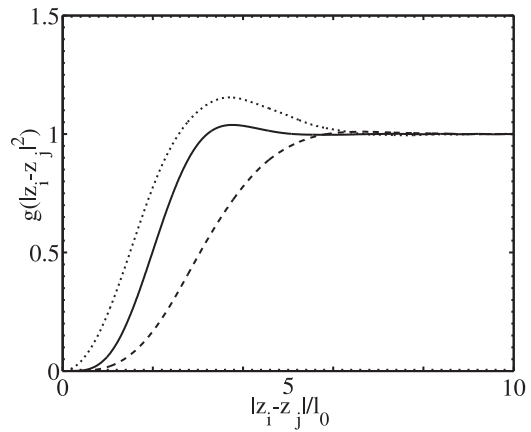}
\caption{Two-point functions: $\nu=1/2$ Laughlin state (solid);
$\tilde\nu=2/3$ 221 state $g_{11}=g_{22}$ (dashed) and
$g_{12}=g_{21}$ (dotted).} \label{fig:correlator}
\end{figure}

We first study how shot noise measurements of the density
correlations \cite{noisecorr-th,bec-lll-expansion} can reveal the
two particle correlation functions $g(|w|^2)$ with $w=x+iy$ and thus
distinguish the FQH states from Mott insulating (MI) states for
which such measurements were recently carried out
\cite{noisecorr-expt}. As shown in \cite{noisecorr-th} results of
these measurements are directly related to the expectation values
\begin{eqnarray}
\av{a^\dagger_i a^\dagger_j a_{j^\prime}
a_{i^\prime}}&=&\sum\limits_{k_i,k_j=1}^{n}d^2{\bf B}^{(k_i)}_i{\bf
B}^{*(k_j)}_j{\bf B}^{*(k_i)}_{i^\prime}{\bf B}^{(k_j)}_{j^\prime}
\nonumber \\
&& \qquad \rho_2(i,j;i^\prime,j^\prime)_{k_ik_j;k_ik_j},
\end{eqnarray}
where $\rho_2(i,j;j^\prime,i^\prime)_{k_ik_j;k_ik_j}$ is the
continuum 2-particle density matrix. For LLL constant density states
$\rho_2(i,j;i^\prime,j^\prime)_{k_ik_j;k_ik_j} \propto
g_{k_ik_j}[(w_i-w_j)^*(w_{i^\prime}-w_{j^\prime})]$
\cite{lll-density-matrix} and is
thus determined by analytic continuation of the two-point
correlation functions $g_{k_ik_j}(|w_i-w_j|^2)$. We have calculated
$g_{k_ik_j}$ for the Laughlin and 221 states by Monte Carlo methods
\cite{lll-density-matrix}. The results in Fig. \ref{fig:correlator}
show clear spatial antibunching up to distances of order $\tilde
\ell_0$ which is very distinct from behavior of superfluid and MI
states.

\begin{figure}
\includegraphics[width=7cm]{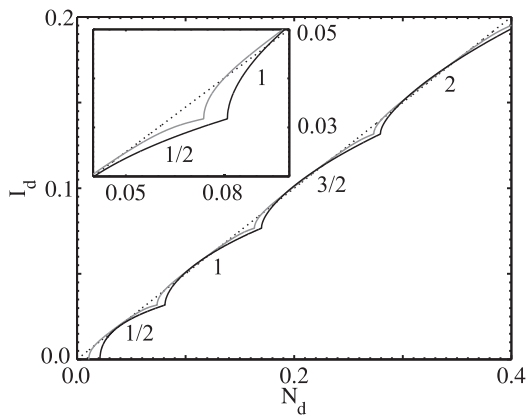}
\caption{Dimensionless Hall current
$I_d=I/(2^{3/2}ma\sqrt{u\Omega}/\omega)$ against dimensionless
number of atoms per unit length
$N_d=N/(2^{3/2}L\Omega^{3/2}m\sqrt{u}/\omega)$ for
$V_1(y)=m\omega^2y^2/2$, $V_2(y)=V_1(y)-may$ at small $\alpha$ and
$a$, for no disorder (dotted straight line), maximal disorder
(constant density of states) (black curve), and Lorentzian disorder
of width $um\Omega/10\pi$ (grey curve). Numbers under curve are
filling factors at trap centre.  The inset shows the $\nu=1/2$ to
$\nu=1$ corner.} \label{fig:current}
\end{figure}

Finally we calculate the Hall currents which can be used to
characterize FQH states near $\alpha_c$. Unlike in a
Galilean-invariant continuum FQH system the application of a linear
potential $V(x,y)=-may$ does not induce a particle current with
velocity of exactly $v_x=a/2\Omega$ since the lattice defines a rest
frame. Instead, when Eq.~(\ref{eq:chi-manypar}) is valid, the
particle velocity is given by $\tilde v_x = 2\tilde\Omega a/(4
\tilde\Omega^2+\omega^2/C)$ if the linear potential is superimposed
to a 1D harmonic trap. The Hall current will thus cease for
$\alpha=\alpha_c$ and for $\alpha \lesssim \alpha_c$ a
counterintuitive negative Hall current flows. In a trapped system at
equilibrium these currents will not be visible since they flow along
equipotentials and steps in the density profile also lie along
equipotentials. They can be made visible by putting the system out
of equilibrium, e.g.~by suddenly changing the trap from $V_1$ to
$V_2$ as shown in Fig. \ref{fig:current}.

When sufficiently mild disorder is added to the optical lattice
\cite{disorder}, some of the particles become localized and cannot
carry current, but the average velocity is still $\tilde v_x$
\cite{fqh-text}. A simple model of this, valid for weak potentials,
is to describe the disorder by a density of states $\rho_d(E)$
giving the number of LLL states at energy $E$ per unit energy
interval per unit area and hence $\int\rho_d(E){\rm
d}E=m\tilde\Omega/\pi$. For a linear geometry we have
\begin{eqnarray}
I&=&\frac{m}{2\pi}\int {\rm
d}y\frac{dV_2}{dy}\int\limits^{\mu-V_1(y)}{\rm d}E\sum\limits_{j\geq
0}(\tilde\nu_{j-1}-\tilde\nu_{j})\delta(E-E_j) \nonumber \\
N&=&L\int {\rm d}y\int\limits^{\mu-V_1(y)}{\rm d}E\sum\limits_{j\geq
0} (\tilde\nu_j-\tilde\nu_{j-1})\rho_d(E-E_j),\label{eq:disorder}
\end{eqnarray}
with $L$ the system's length, $I$ the net current (both in the
$x$-direction) and $\mu$ the chemical potential. The FQH states are
at filling factors $\tilde\nu_j$ and energies per particle $E_j$
[for $\alpha\approx 0$, $\nu_j=(j+1)/2$ and $E_j\approx
(um\Omega/2\pi)j$].

In a harmonic trap the disorder will not lead to FQH plateaus, only
corners each time a new extended level begins to fill
(c.f.~Fig.~\ref{fig:current}), but unlike the square-well case, it
is possible to obtain the complete distribution $\rho_d(E)$ by
measuring $I$ against $N$ (or $\Omega$) and using
Eqs.~(\ref{eq:disorder}). The quasiparticles of our system carry
fractional particle number \cite{fqh-current-noise} and fractional
statistics \cite{bec-anyons} as in the conventional FQH system, but
detecting this directly is likely to be more experimentally
challenging.

In summary we have investigated bosonic optical lattice FQH systems
at large magnetic fields. We have worked out their ground states,
magnetic lengths and the resulting fractional fillings, and found
the Hall current to change its sign at $\alpha_c$. Furthermore, we
have studied ways for detecting these properties using recently
developed experimental methods. This work is supported by EPSRC
through project EP/C51933/1 and QIP IRC (GR/S82176/01), and by the
EU through OLAQUI.

\end{document}